\documentclass[times]{article}
\usepackage[noadjust]{cite}

\usepackage{isita2008}

  \usepackage[dvips]{graphicx}
  \graphicspath{{./figures/}}

\usepackage[cmex10]{amsmath}
\usepackage{amssymb}
\usepackage{amsfonts}
\usepackage{algorithmic}

\newtheorem{theorem}{Theorem}
\newtheorem{conjecture}{Conjecture}
\newtheorem{definition}{Definition}

\newtheorem{remark}{Remark}

\newcommand{\ie}{{\it i.e.}}

\title{Online network coding for optimal throughput and delay -- the three-receiver case}
\name{Jay~Kumar~Sundararajan,
        Devavrat~Shah,
        Muriel~M\'edard\vspace{-.7in}\thanks{The authors are in the Department of Electrical Engineering and Computer Science, at the Massachusetts Institute of Technology, Cambridge, MA 02139. Email: \{jaykumar, devavrat, medard\}@mit.edu. This material is based upon work under subcontract \# 18870740-37362-C issued by Stanford University and supported by DARPA, and upon work under a subcontract \# 060786 issued by BAE Systems National Security Solutions, Inc. and supported by DARPA and the Space and Naval Warfare System Center (SPAWARSYSCEN), San Diego under Contract No. N66001-06-C-2020.}
}

\begin{document}
\maketitle

\begin{abstract}
For a packet erasure broadcast channel with three receivers, we propose a new coding algorithm that makes use of feedback to dynamically adapt the code. Our algorithm is throughput optimal, and we conjecture that it also achieves an asymptotically optimal average decoding delay at the receivers. We consider heavy traffic asymptotics, where the load factor $\rho$ approaches 1 from below with either the arrival rate ($\lambda$) or the channel parameter ($\mu$) being fixed at a number less than 1. We verify through simulations that our algorithm achieves an asymptotically optimal decoding delay of $O\left(\frac1{1-\rho}\right)$. 
\end{abstract}

\section{Introduction}
Reliable communication over packet erasure channels is a well studied problem. Several solutions have been proposed, each with its own requirements, merits and issues. In this work, we consider a three-receiver packet erasure broadcast channel with feedback and address questions of throughput and decoding delay at the receivers. 

To communicate over a packet erasure broadcast channel, one can use the random linear network coding solution of \cite{desmondallerton}, where the sender transmits a random linear combination of all packets that have arrived so far. Digital fountain codes (\cite{ltcodes,raptor}) form another approach to this problem. These solutions use coding to ensure that with high probability, the transmitted packet will have what we call the \emph{\bf innovation guarantee property}, \ie, it will be \emph{innovative}\footnote{An innovative packet is a coded packet whose coefficient vector is outside the span of previously received packets' coefficient vectors.} to every receiver that receives it, except if the receiver already knows as much as the sender. Thus, every successful reception brings a unit of new information. Such schemes achieve 100\% throughput.

However, both fountain codes and random linear network coding perform block-based encoding. In general, the receiver may not be able to extract the original packets from the coded packets till the entire block has been received. This leads to a decoding delay, which is a problem for real-time packet streaming applications such as video. Ideally we want a code that would allow playback even before the full block is received. In other words, we are interested in minimizing the average per-packet delay. Related questions have been studied by \cite{eminthesis}, \cite{sujay}, \cite{rtoblivious} and \cite{keller}. 

With full feedback, the optimal scheme over a point-to-point packet erasure channel is Automatic Repeat reQuest (ARQ) -- the sender simply retransmits a packet upon erasure. This scheme also has the advantage of being a sliding window approach as opposed to a block-based approach. Although it achieves 100\% throughput and optimal packet delay, ARQ does not extend to broadcast-mode links. On the other hand, network coding readily extends to broadcast-mode links.

Reference \cite{ARQISIT} proposes a scheme that uses feedback to acknowledge degrees of freedom instead of original packets, thus combining the benefits of network coding and ARQ. This new framework allows the sender to dynamically adapt its code to incorporate receivers' states of knowledge. This fact was used to design a queue management algorithm called the drop-when-seen algorithm that minimizes the sender's queue size, along with a coding module that provides 100\% throughput. Another related reference is \cite{tran}, where the authors combine an acknowledgment scheme with network coding. Here, the main focus is to maximize the throughput. In contrast to these works, our current work focuses on achieiving the best possible decoding delay for all receivers, while maintaining optimal throughput. 

By decoding delay of a receiver, we mean the time elapsed between the arrival of a packet into the sender's queue and its getting decoded by the receiver under consideration, averaged over the packets in the long run in a packet streaming scenario. This is different from but related to the notion of delay used in \cite{keller}. 

For the special case of a packet erasure broadcast channel with only two receivers, reference \cite{durvy} proposes a feedback-based coding algorithm that not only achieves 100\% throughput, but also guarantees that every successful innovative reception will cause the receiver to decode a new packet. We call this property \emph{instantaneous decodability}. Instantaneous decodability and 100\% throughput are both desirable goals. However, this approach does not extend to the case of more than two receivers. With prior knowledge of the erasure pattern, \cite{keller} gives an offline algorithm that achieves optimal delay and throughput for the case of three receivers. However, in the online case, even with only three receivers, \cite{durvy} shows through an example that it is not possible to simultaneously guarantee instantaneous decodability as well as throughput optimality. 

In the light of this example, our current work aims for a relaxed version of instantaneous decodability while still retaining the requirement of optimal throughput. Our relaxation of the condition is as follows. Let $\lambda$ and $\mu$ be the arrival rate and the channel quality parameter respectively. Let $\rho\triangleq \lambda/\mu$ be the load factor. We consider asymptotics when the load factor on the system tends to 1 (\ie, 100\%), while keeping either $\lambda$ or $\mu$ fixed at a number less than 1. The optimal throughput requirement means that the queue of undelivered packets is stable for all values of $\rho$ less than 1. Our new requirement on decoding delay is that the growth of the average decoding delay as $\rho\rightarrow 1$ should be the same as for the single receiver case. The expected per-packet delay of a receiver in a system with more than one receiver is clearly lower bounded by the corresponding quantity for a single-receiver system. Thus, instead of optimal decoding delay, we aim to guarantee asymptotically optimal decoding delay. The motivation is that in most practical systems, delay becomes a critical issue only when the system starts approaching its full capacity. When the load on the system is well within its capacity, the delay is usually small and hence not an issue. For the case of two receivers, it can be shown that this relaxed requirement is satisfied by the scheme in \cite{durvy} due to the instantaneous decodability property, \ie, the scheme achieves the asymptotically optimal average decoding delay per packet for the two-receiver case. 

In our current work, we provide a new coding module for the case of three receivers that achieves optimal throughput. We conjecture that at the same time it also achieves an asymptotically optimal decoding delay in the following sense. With a single receiver, the optimal scheme is ARQ with no coding and we show that this achieves an expected per-packet delay at the sender of $\Theta\left(\frac{1}{1-\rho}\right)$. For the three-receiver system, we conjecture that our scheme also achieves a delay of $O\left(\frac{1}{1-\rho}\right)$, and thus meets the lower bound in an asymptotic sense. We have verified this behavior through simulations for values of $\rho$ that are very close to 1. Our scheme thus achieves feedback-based control of the decoding delay, along the lines suggested in \cite{desmondfeedback}. We believe that our approach can be extended to an arbitrary number of receivers as well.

\section{Preliminaries}

\subsection{The setup}
\begin{figure}
	\centering
		\includegraphics[width=0.45\textwidth]{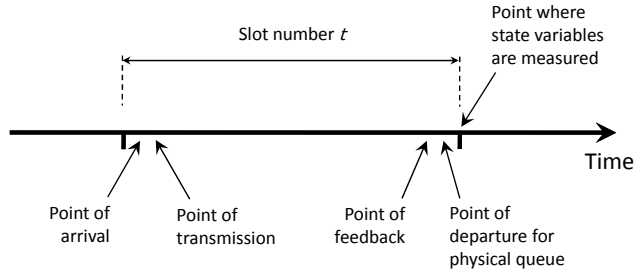}
		\caption{Relative timing of arrival, service and departure points within a slot}\label{earlyarrival}
\end{figure}

The setup is the same as in \cite{ARQISIT}. Time is slotted. Packets arrive into the sender's queue according to a Bernoulli process of rate $\lambda$. The sender wants to broadcast this stream to three receivers over a packet erasure broadcast channel. The sender has one queue with no preset size constraints. The channel accepts one packet per slot. Each receiver either receives this packet with no errors (with probability $\mu$) or an erasure occurs (with probability $(1-\mu)$). Erasures occur independently across receivers and across slots and can be detected by receivers. There is perfect feedback in each slot. Figure \ref{earlyarrival} shows the timing of events in a slot. In particular, we assume that the sender finds out whether the receivers received the previous slot's transmission before selecting the current slot's transmission. 

\subsection{The lower bound}\label{lowerbound}
\begin{figure}
	\centering
		\includegraphics[width=0.45\textwidth]{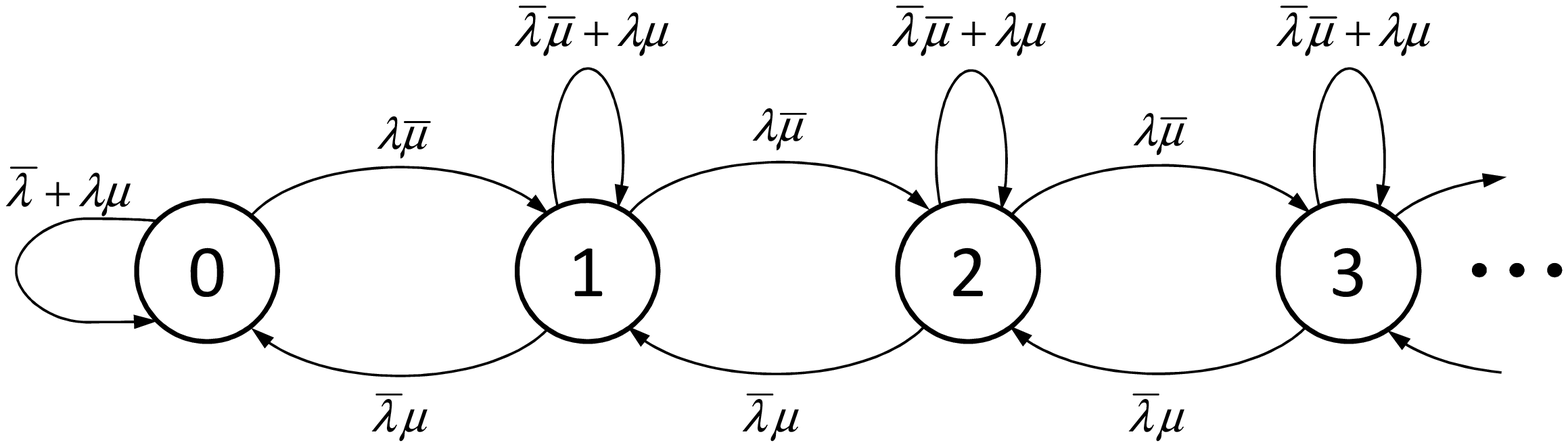}
		\caption{Markov chain for the sender's queue size -- single receiver case. Here $\bar{\lambda}:=(1-\lambda)$ and $\bar{\mu}:=(1-\mu)$. }\label{markovchain}
\end{figure}

The expected per-packet delay for the single receiver case is clearly a lower bound for the corresponding quantity at one of the receivers in a three-receiver system. We will compute this lower bound in this section. Figure \ref{markovchain} shows the Markov chain for the queue size. If $\rho:=\frac\lambda\mu<1$, then the chain is positive recurrent and its steady state distribution can be found (\cite{hunterbook}). Based on this, the steady state expected queue size can be computed to be $\frac{\rho(1-\mu)}{(1-\rho)}=\Theta\left(\frac1{1-\rho}\right)$. Now, if $\rho<1$, then the system is stable and Little's law can be applied to show that the expected per-packet delay in the single receiver system is also $\Theta\left(\frac1{1-\rho}\right)$. 

\subsection{Representing knowledge}
We treat packets as vectors over some finite field. Throughout this paper, we consider a single source that generates a stream of packets. The $k^{th}$ packet that the source generates is said to have an \emph{index} $k$ and is denoted as $\mathbf{p_k}$. We assume that the sender uses only linear codes, \ie, the transmission is some linear combination of packets. The linear combination is uniquely specified using the vector of coefficients used to form it.
With linear codes, the state of knowledge of a node after receiving some set of linear combinations has a vector space structure. This is because the node can compute any linear combination whose coefficient vector is within the linear span of the coefficient vectors of previously received linear combinations. This leads to the following definition of \emph{knowledge space} which we restate from \cite{ARQISIT}. 

\begin{definition}[Knowledge of a node]
	The \emph{knowledge of a node} is the set of all linear combinations of original packets that it can compute, based on the information it has received so far. The coefficient vectors of these linear combinations form a vector space called the \emph{knowledge space} of the node. The dimension of this vector space is called the \emph{rank} of the node. 
\end{definition}

We next restate the definition of a node ``seeing'' a message packet from \cite{ARQISIT}. A node is said to have \emph{seen} a message packet $\mathbf{p}$ if it has received enough information to compute a linear combination of the form $(\mathbf{p} + \mathbf{q})$, where $\mathbf{q}$ is itself a linear combination involving only packets with an index greater than that of $\mathbf{p}$. (Decoding implies seeing, as we can pick $\mathbf{q}=\mathbf{0}$.) The number of packets seen by a node is precisely the rank of the node (see \cite{ARQISIT} for more details).

We introduce a new notion of packets that a node has ``heard of''. A node is said to have heard of a packet if it knows some linear combination involving that packet.

\section{The new coding module}
Our coding module works in the Galois field of size 3. At the beginning of every slot, the module has to decide what linear combination to transmit. Since there is full feedback, the module is fully aware of the current state of knowledge of each of the three receivers. Thus, it can compute the rank of each receiver. We denote the highest rank among the three receivers as $m$. Our coding module maintains an invariant that the transmission will never involve a packet whose index is greater than $(m+1)$. We denote the receiver(s) whose rank is $m$ as the leader(s). We consider three cases:
\subsection{All three receivers are leaders}
In this case, all three receivers have a rank of $m$ which means each has seen $m$ packets. If $\mathbf{p_{m+1}}$ has not arrived yet, the module does nothing. Otherwise, since all transmissions have involved only the set of packets up to $\mathbf{p_{m+1}}$, there is exactly one unseen packet for each receiver within this set. This could be a different packet for each of the three receivers. The coding module selects a linear combination that if received successfully by any receiver, will reveal to that receiver its unseen packet, thereby guaranteeing innovation. This is done by simply forming a linear combination involving only the unseen packets of the three receivers. It can be verified that with a field of size 3, it is always possible to choose coefficients such that innovation is guaranteed for all three receivers. 

\subsection{There are two leaders}
It can be shown by induction that at all times, at least one leader would have decoded all packets from 1 to $m$. Now, when there are two leaders, if exactly one leader has decoded all packets 1 to $m$, then the coding module performs the operations of case 3, treating this leader as the unique leader. If both leaders have decoded packets 1 to $m$, then module does the following.

If $\mathbf{p_{m+1}}$ has not arrived yet, the module transmits the oldest undecoded packet of the non-leader (if there are packets that the non-leader has heard of but not yet decoded, then they are preferred). Suppose $\mathbf{p_{m+1}}$ has arrived. Now, if it has already been decoded by the non-leader, then the module sends the sum of $\mathbf{p_{m+1}}$ and the oldest undecoded packet of the non-leader (again, if there are packets that the non-leader has heard of but not yet decoded, then they are preferred). Otherwise, the module sends $\mathbf{p_{m+1}}$ by itself. 

\subsection{Unique leader}
In this case, the module computes the following sets for the two non-leaders:

$H_1$:= Set of packets heard of by first non-leader

$H_2$:= Set of packets heard of by second non-leader

$D_1$:= Set of packets decoded by first non-leader

$D_2$:= Set of packets decoded by second non-leader

\begin{figure}
	\centering
		\includegraphics[width=0.45\textwidth]{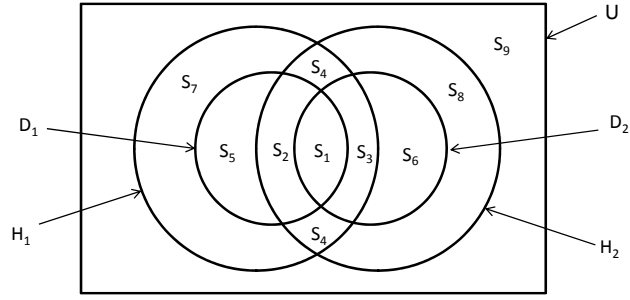}
		\caption{Sets used by the coding module}\label{venn}
\end{figure}

Note that $D_1\subseteq H_1$ and $D_2\subseteq H_2$. We also define a universe set $U$ consisting of packets $\mathbf{p_1}$ to $\mathbf{p_m}$, and also $\mathbf{p_{m+1}}$ if it has arrived. In this setting, the following sets partition the universe (refer to Figure \ref{venn}):
\begin{itemize}
\item $S_1=D_1\cap D_2$
\item $S_2=D_1\cap(H_2\backslash D_2)$
\item $S_3=D_2\cap(H_1\backslash D_1)$
\item $S_4=(H_1\backslash D_1)\cap (H_2\backslash D_2)$
\item $S_5=D_1\backslash H_2$
\item $S_6=D_2\backslash H_1$
\item $S_7=(H_1\backslash D_1)\backslash H_2$
\item $S_8=(H_2\backslash D_2)\backslash H_1$
\item $S_9=U\backslash(H_1\cup H_2)$
\end{itemize}

The coding module picks a linear combination depending on which of these sets $\mathbf{p_{m+1}}$ falls in, as follows:

{\it Case 1 -- $\mathbf{p_{m+1}}$ has not arrived:} 
Check if $S_4$ is non-empty. If it is, then send the oldest packet in $S_4$. Otherwise, check if both $S_2$ and $S_3$ are non-empty. If they are, pick the oldest packet from each, and send their sum. If not, try the following pairs of sets: $S_3$ and $S_5$, else $S_2$ and $S_6$, else $S_5$ and $S_6$. If none of these pairs of sets work, then send the oldest packet in $S_7$ if it is non-empty. If not, try $S_8$, $S_9$, $S_2$, $S_3$, $S_5$ and $S_6$ in that order. If all of these are empty, then send nothing.

{\it Case 2 -- $\mathbf{p_{m+1}}\in S_1$:} This is identical to case 1, except that $\mathbf{p_{m+1}}$ must also be added to the linear combination that case 1 suggests.

{\it Case 3 -- $\mathbf{p_{m+1}}\in S_2$:} Send $\mathbf{p_{m+1}}$ added to another packet. The other packet is chosen to be the oldest packet in the first non-empty set in the following sets, tested in this particular order: $S_3$, $S_4$, $S_6$, $S_8$, $S_7$ and then $S_9$. 

{\it Case 4 -- $\mathbf{p_{m+1}}\in S_3$:} This is similar to the $S_2$ case (using symmetry) -- test $S_2$, $S_4$, $S_5$, $S_7$, $S_8$ and then $S_9$.

{\it Case 5 -- $\mathbf{p_{m+1}}\in S_4$:} Send $\mathbf{p_{m+1}}$ as it is.

{\it Case 6 -- $\mathbf{p_{m+1}}\in S_5$:} Send $\mathbf{p_{m+1}}$ added to another packet. The other packet is chosen to be the oldest packet in the first non-empty set in the following sets, tested in the following order: $S_3$, $S_6$, $S_4$, $S_8$, $S_7$ and then $S_9$.

{\it Case 7 -- $\mathbf{p_{m+1}}\in S_6$:} This is similar to the $S_5$ case (using symmetry) -- test $S_2$, $S_5$, $S_4$, $S_7$, $S_8$ and then $S_9$.

{\it Case 8 -- $\mathbf{p_{m+1}}\in S_7$:} Send $\mathbf{p_{m+1}}$ as it is.

{\it Case 9 -- $\mathbf{p_{m+1}}\in S_8$:} Send $\mathbf{p_{m+1}}$ as it is.

{\it Case 10 -- $\mathbf{p_{m+1}}\in S_9$:} Send $\mathbf{p_{m+1}}$ as it is.

In all these cases, the coefficient for the chosen packets must be selected to be either 1 or 2, in such a way that the resulting linear combination is innovative to any receiver that receives it, except if the receiver already knows all that the sender knows. It can be shown that such a choice is always possible with a field of size 3.

\begin{remark}
\emph{We conjecture based on the simulations that the algorithm maintains the following invariant -- at most one of $H_1\backslash D_1$ and $H_2\backslash D_2$ is non-empty at any given time. If proved to be true, this observation can be used to simplify the algorithm.}
\end{remark}

\section{The intuition behind the algorithm}
The main idea behind the algorithm is to first of all guarantee innovation. It can be shown that the linear combination computed by this coding module is indeed innovative to any receiver that receives it. In addition to this requirement however, the module also tries to cause each receiver that has a successful reception to decode as many packets as possible. 

An interesting property of this algorithm is that the transmitted linear combination always has at most two undecoded packets involved in it from any receiver's point of view. In other words, every transmission is essentially either an uncoded packet or the sum of two packets. This property leads to a nice structure in the knowledge space of the receivers, using which, we present a strategy to control the extent to which packets get mixed with each other, thereby controlling the decoding delay.

In order to explain this structure, we define the following relation. The ground set $G$ of the relation contains all packets that have arrived at the sender so far, along with a fictitious all-zero packet that is known to all receivers even before transmission begins. The relation is defined with respect to a specific receiver. Two packets $\mathbf{p_x}\in G$ and $\mathbf{p_y}\in G$ are defined to be related to each other if the receiver knows at least one of $\mathbf{p_x}+\mathbf{p_y}$ and $\mathbf{p_x}+2\mathbf{p_y}$.

Now, a packet added with two times the same packet gives $\mathbf{0}$ which is trivially known to the receiver. Hence, the relation is reflexive. It is also symmetric since addition is a commutative operation. Now, for any $\mathbf{p_x}, \mathbf{p_y}, \mathbf{p_z}$ in $G$, if a receiver knows $\mathbf{p_x}+ \alpha\mathbf{p_y}$ and $\mathbf{p_y}+ \beta\mathbf{p_z}$, then it can compute either  $\mathbf{p_x}+\mathbf{p_z}$ or $\mathbf{p_x}+2\mathbf{p_z}$ by canceling out the $\mathbf{p_y}$, for $\alpha=1$ or 2 and $\beta=1$ or 2. Therefore the relation is also transitive and is thus an equivalence relation. It defines a partition on the ground set, namely the equivalence classes, which provide a structured way to represent the knowledge of the node. It can be seen that the class containing the all-zero packet is precisely the set of decoded packets ($D_1$ or $D_2$). Packets that have not been involved in any of the successfully received linear combinations so far will form singleton equivalence classes. These correspond to the packets that the receiver has not heard of $(U\backslash H_1$ or $U\backslash H_2$). 

We say a class is nontrivial if it is neither a singleton class nor the class of decoded packets. Thus, nontrivial classes contain the packets that have been heard of but not decoded. Revealing any packet in a class will reveal the entire class to the node. The number of nontrivial classes is thus the number of packets that the node needs to know in order to decode all packets it has heard of. This number is thus a measure of how far away a node is from decoding all packets it has heard of. We call this number the \emph{deficit of the node}.

For instance, revealing a packet from $H_1\backslash D_1$ will allow the entire class containing that packet to be decoded by receiver 1. The algorithm ensures that a packet from $H_1\backslash D_1$ or $H_2\backslash D_2$ is revealed whenever possible, as opposed to a packet that the receiver has not heard of. This ensures that the deficit is reduced whenever possible. As a result, the deficit drops to zero frequently, thereby causing the node to decode packets. 

\section{Performance of the algorithm}
\subsection{Throughput optimality}
The algorithm has been designed in such a way that innovation is guaranteed to all the receivers whenever possible. Packet $\mathbf{p_{m+1}}$ is always included in the linear combination if it has arrived, in order to guarantee innovation to the leader. If both the other receivers have also not decoded it, then sending $\mathbf{p_{m+1}}$ by itself satisfies the innovation guarantee. This happens in cases 5, 8, 9 and 10. 

If however, some receiver has already decoded it as in the other cases, then another packet is included in the linear combination that the receiver has not yet decoded, thereby ensuring innovation. While choosing such a packet, preference is given to packets that the receiver has heard of, as revealing such a packet will cause several packets to be decoded at once. 

If $\mathbf{p_{m+1}}$ has not yet arrived, then the leader is already satisfied. For the other two receivers, the transmission is selected in such a way that it simultaneously reveals an undecoded packet to both of them whenever possible. 

We can show that in all these cases, over a field of size 3, the coefficients can also be chosen carefully to guarantee innovation for all those who receive the linear combination successfully. This discussion is summarized in the following theorem. 
\begin{theorem}\label{innovation_for_modified}
\it The coding module satisfies the innovation guarantee property. 
\end{theorem}
This means that the algorithm achieves optimal throughput, \ie, for all $\rho<1$, the decoding delay and the queue at the sender will remain stable.

\subsection{Decoding delay}
\begin{figure}
  \begin{center}
	\includegraphics[width=0.5\textwidth]{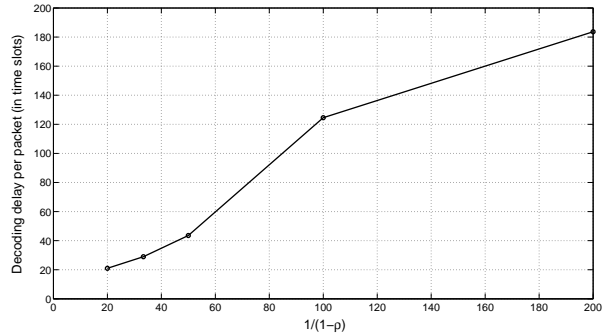}
  \end{center}
  \caption{Decoding delay as load approaches capacity}
  \label{normal}
\end{figure}

We now study the delay experienced by an arbitrary arrival before it gets decoded by one of the receivers, say receiver 1. We consider a system where $\mu$ is fixed at 0.5. The value of $\rho$ is varied as follows: 0.95, 0.97, 0.98, 0.99 and 0.995. We plot the expected decoding delay per packet averaged across the three receivers, as a function of $\left(\frac{1}{1-\rho}\right)$ in Figure \ref{normal}. We also plot the log of the same quantities in Figure \ref{loglog}. The value of the delay is averaged over $500000$ time slots for the first three points and $10^6$ time slots for the last two points.

Figure \ref{normal} shows that the growth of the expected decoding delay is linear in $\left(\frac{1}{1-\rho}\right)$ as $\rho$ approaches 1. Figure \ref{loglog} confirms this behavior -- we can see that the slope on the plot of the logarithm of these quantities is indeed close to 1. This observation leads to the following conjecture:

\begin{figure}[t!]
  \begin{center}
	\includegraphics[width=0.5\textwidth]{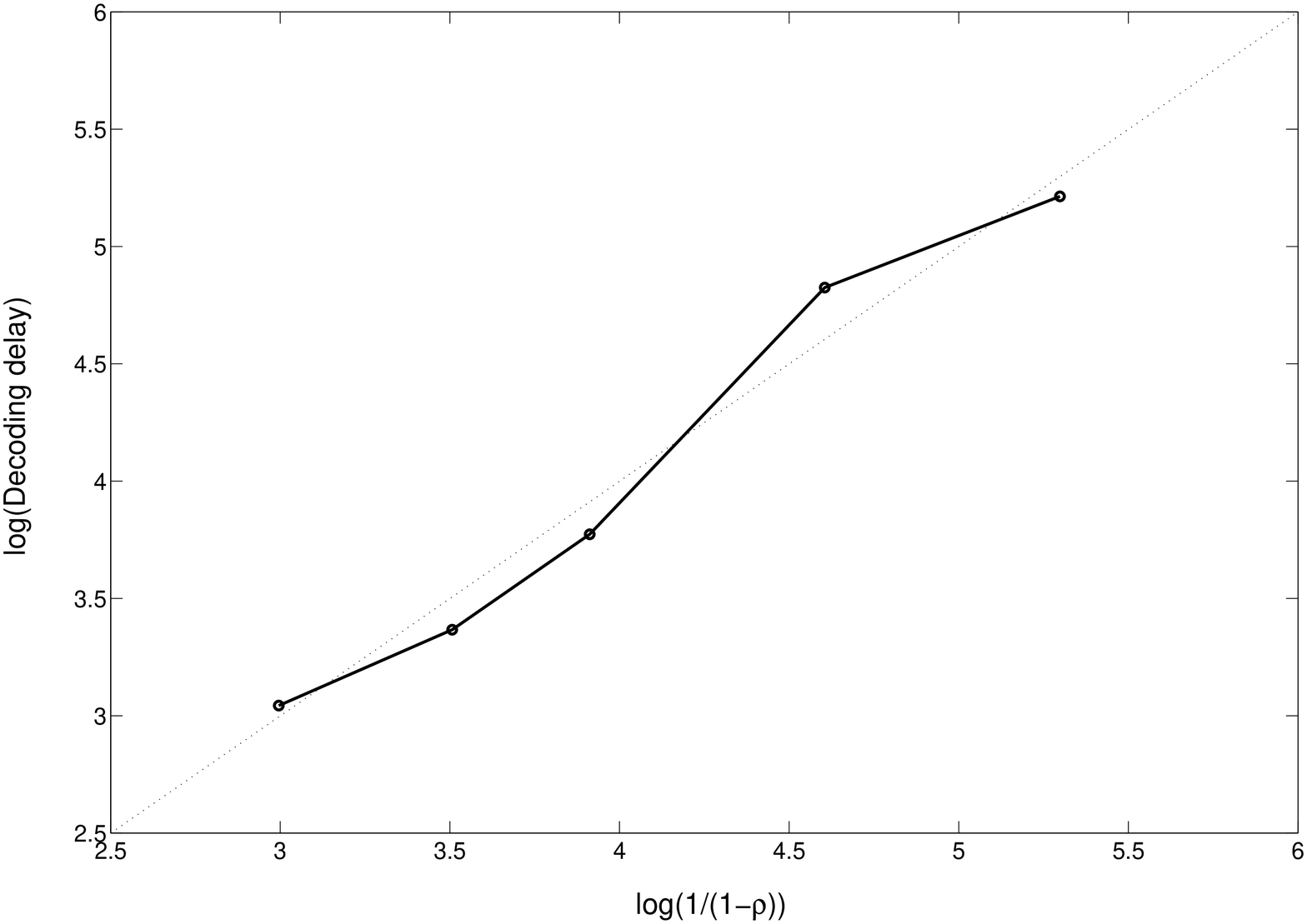}
  \end{center}
  \caption{Decoding delay -- log-log plot}
  \label{loglog}
\end{figure}

\begin{conjecture}\label{conj1}
For the newly proposed coding module, the expected decoding delay per packet from one particular receiver's point of view grows as $O\left(\frac{1}{1-\rho}\right)$, which is asymptotically optimal.
\end{conjecture}

\subsection{Queue size}
The queue update rule is as follows -- the sender drops a packet if all the receivers have decoded it. This means that by Little's law, the expected queue size will be proportional to the time a packet spends in the system before it is decoded. Thus, if the expected decoding delay is indeed $O\left(\frac1{1-\rho}\right)$ as conjectured above, then the drop-when-decoded queue update rule will ensure that the expected queue size at the sender is $O\left(\frac1{1-\rho}\right)$, which is asymptotically optimal. 

\section{Conclusions}\label{conc}
For a three receiver packet erasure broadcast channel with feedback, we have proposed a new coding scheme that makes use of the feedback to dynamically adapt the code. As argued earlier, $\Theta\left(\frac1{1-\rho}\right)$ is an asymptotic lower bound on the decoding delay. We have observed through simulations that this lower bound seems to be achieved by our scheme, which would imply the asymptotic optimality of our coding module in terms of decoding delay. We conjecture that this is indeed true. All these delay benefits are obtained without compromising on throughput. If the conjecture is true, then the expected queue size of undecoded packets is also $O\left(\frac1{1-\rho}\right)$, which is asymptotically optimal. Thus, our scheme also simplifies the queue management at the sender. In the future, we wish to extend this approach to an arbitrary number of receivers. Also, we wish to make the algorithm robust to delays and erasures in the feedback. 
\bibliographystyle{IEEEtran}
\bibliography{References}
\end{document}